\documentclass[aps,prb,twocolumn,showpacs,preprintnumbers,amsmath,amssymb,superscriptaddress,longbibliography,nofootinbib]{revtex4-2}
\usepackage{hyperref}
\hypersetup{colorlinks=true, citecolor=blue, urlcolor=blue, linkcolor=blue,urlcolor=cyan}
\usepackage{color}
\usepackage{graphicx}
\usepackage{dcolumn}
\usepackage{braket}
\usepackage{amsthm}
\usepackage[caption=false]{subfig}

\newcommand{\ex}[1]{\mathrm{e}^{#1}}

\newcommand{\pa}[1]{\left(#1 \right)}

\newcommand{\ca}[1]{\mathcal{#1}}

\newcommand{\abs}[1]{\left|#1\right|}

\newcommand{\kett}[1]{ \ket{#1} \rangle }

\newcommand{\fr}{\frac}

\def\be{\begin{equation}}
\def\ee{\end{equation}}
\def\ba{\begin{eqnarray}}
\def\ea{\end{eqnarray}}

 \def\ba{{\bar{\alpha}}}

\def\tr{{\text{tr}}}

\begin{document}

\title{Universal Bound on Effective Central Charge and Its Saturation}
\preprint{CALT-TH 2024-014}
\preprint{IPMU 24-0010}
\preprint{RIKEN-iTHEMS-Report-24}
\preprint{KYUSHU-HET-285}

\author{Andreas Karch}
\affiliation{\it Theory Group, Weinberg Institute, Department of Physics, University of Texas, 2515 Speedway, Austin, TX 78712-1192, USA}

\author{Yuya Kusuki}
\affiliation{\it Walter Burke Institute for Theoretical Physics, California Institute of Technology, Pasadena, CA 91125, USA}
\affiliation{\it Interdisciplinary Theoretical and Mathematical Sciences (iTHEMS), RIKEN, Wako, Saitama 351-0198, Japan}
\affiliation{\it Institute for Advanced Study, Kyushu University, Fukuoka 812-8581, Japan}
\affiliation{\it Department of Physics, Kyushu University, Fukuoka 819-0395, Japan}

\author{Hirosi Ooguri}
\affiliation{\it Walter Burke Institute for Theoretical Physics, California Institute of Technology, Pasadena, CA 91125, USA}
\affiliation{\it Kavli Institute for the Physics and Mathematics of the Universe (WPI), University of Tokyo, Kashiwa 277-8583, Japan}

\author{Hao-Yu Sun}
\affiliation{\it Theory Group, Weinberg Institute, Department of Physics, University of Texas, 2515 Speedway, Austin, TX 78712-1192, USA}

\author{Mianqi Wang}
\affiliation{\it Theory Group, Weinberg Institute, Department of Physics, University of Texas, 2515 Speedway, Austin, TX 78712-1192, USA}

\begin{abstract}

The effective central charge (denoted by $c_{\text{eff}}$) is a measure of entanglement through a conformal interface, while the transmission coefficient (encoded in the coefficient $c_{LR}$ of the two-point function of the energy-momentum tensor across the interface) is a measure of energy transmission through the interface. It has been pointed out that these two are generally different. In this article, we propose the inequalities, $0 \leq c_{LR} \leq c_{\text{eff}} \leq \min (c_L,c_R)$. They have the simple but important implication that the amount of energy transmission can never exceed the amount of information transmission. We verify them using the AdS/CFT correspondence, using the perturbation method, and in examples beyond holography. 
We also show that these inequalities are sharp by constructing a class of interfaces that saturate them.

\end{abstract}

\maketitle

\section{Introduction \& Summary}

Conformal interfaces play an important role in the study of quantum critical systems.  However, our knowledge of their general properties is limited because they break half of the conformal symmetry. The AdS/CFT correspondence is useful in this context because it gives us insight into systems far from free fields. Indeed, 
our recent work \cite{Karch2023} demonstrated a utility of the AdS/CFT correspondence in studying general properties of interfaces.  In this paper we extend this line of research and formulate the conjecture that the effective central charge, which measures the entanglement across a conformal interface in $1+1$ dimensions, is bounded below by the two-point function of the energy-momentum tensor across the interface. Namely, the amount of energy transmitted across the interface cannot exceed the amount of information transmitted.
The conjecture is motivated by holographic CFTs, free field examples, and the defect perturbation theory. We also show that the bound can be saturated by constructing explicit examples. 

In $1+1$ dimensions, conformal interfaces are defined by the following boundary condition for the energy-stress tensors across the interface \cite{Oshikawa1996,Oshikawa1997,Bachas2002}:
\begin{equation}
T^{(L)}-\bar{T}^{(L)} = T^{(R)} - \bar{T}^{(R)},
\end{equation} 
where $T^{(i)}$ and $\bar{T}^{(i)}$ are the holomorphic and anti-holomorphic energy-stress tensors of CFT${}_i$, respectively.
In the operator formalism, this can be re-expressed using the Virasoro generators $L_n^{(i)}$ and $\bar{L}_{-n}^{(i)}$ in CFT${}_i$ as
\begin{equation}\label{eq:I}
\pa{L_n^{(L)} - \bar{L}_{-n}^{(L)} }\ca{I} = \ca{I} \pa{L_n^{(R)} - \bar{L}_{-n}^{(R)} }  \ \ \ \ \ \ \forall \,n.
\end{equation}
This condition does not fully determine an interface:
It demands the interface not to absorb energy while allowing flexibility regarding the amount of energy reflected by the interface.

Conformal interfaces can be characterized by the {\it effective central charge}, which controls the amount of entanglement across them.
Entanglement entropy between two (possibly different) systems has the following form,
\begin{equation}
S_A = \fr{c_{\text{eff}}}{3}\ln \fr{L}{\pi \epsilon},
\end{equation}
where the system size for CFT${}_L$ and CFT${}_R$ is denoted as $L$, $\epsilon$ is the lattice regularization parameter, and $c_{\text{eff}}$ is the effective central charge.
A convenient way to evaluate the entanglement entropy is given by the replica trick,
\begin{equation}
S_A = \lim_{n \to 1} S_A^{(n)},
\ \ \ \ \ \ 
S_A^{(n)} = \fr{1}{1-n} \ln \fr{Z_n}{(Z_1)^n},
\end{equation}
where the replica partition function is defined as
\begin{equation}
Z_n \equiv
 \text{tr} \ \pa{\ex{-\fr{\beta}{2} H^{(L)}}
 \ca{I}^{L \to R}
 \ex{-\fr{\beta}{2} H^{(R)}}
\ca{I}^{R \to L}
}^n.
\end{equation}
The dependence on the subsystem size is encoded in the temperature as $\beta = \fr{1}{\pi} \ln \fr{L}{\pi \epsilon} $.
We can define the interface Hilbert space $\mathcal{H}_n^I$ by the dual-channel expansion of the replica partition function,
\begin{equation}
Z_n = \tr_{\mathcal{H}_n^I} \  \ex{ -\fr{(2\pi)^2}{\beta}H^{\ca{I}}_n },
\end{equation}
where we formally define the Hamiltonian $H^{\ca{I}}_n$ in the presence of the interfaces.
Then, one can give an alternative definition of the effective central charge in terms of the vacuum energy $\Delta^0_n$ in the interface Hilbert space as
\begin{equation}\label{eq:ceff}
c_\text{eff} \equiv \lim_{n \to 1} \fr{12n}{1-n^2} \pa{n\Delta^0_1 - \fr{\Delta^0_n}{n}}.
\end{equation}
The effective central charge has been calculated in some specific models \cite{Sakai2008, Brehm2015,Brehm2016,Wen2018,Gutperle2016a}.
Nevertheless, there is still much unknown about $c_\text{eff}$ due to the lack of techniques in interface CFT (ICFT) where the conformal symmetry is partially broken by interfaces.

Another quantity known to characterize interfaces is the {\it transmission coefficient}, which measures the transfer of energy across the interface \cite{Quella2007}. This quantity is controlled by the two-point function of the stress tensor across the interface,
\begin{equation}
\braket{T^{(L)}(z_1) T^{(R)}(z_2)} = \fr{c_{LR}}{2(z_1-z_2)^4}.
\end{equation}
The weighted average transmission coefficient can be expressed in terms of $c_{LR}$ as
\begin{equation}
\ca{T} = \fr{2c_{LR}}{c_L+c_R},
\label{eq:defclr}
\end{equation}
where $c_L$ and $c_R$ are the central charges of the two CFTs connected by the interface.
Similar expressions in terms of $c_{LR}$ can be given for transmission from left and right separately \cite{Meineri2020}.
Here we would like to emphasize that the transmission of energy across the interface is independent of the transmission of information.
One will see this independence later in this article.

One of our main results is to provide evidence for the following inequality,
\begin{equation}
c_{LR} \leq c_{\text{eff}}.
\label{eq:lower}
\end{equation}
It implies that the amount of energy transmitted across the interface cannot exceed the amount of information transmitted, which is directly controlled by $c_{\text{eff}}$ \cite{Wen2018}.
We have confirmed that this inequality holds in general holographic CFTs. Furthermore, it also holds in free CFT beyond holography.
We also verify the inequality in the defect perturbation theory.
Based on these examples, we propose it to hold in general CFTs.

Using the entropic $c$-theorem, it has been shown that there is an upper bound on $c_{\text{eff}}$ \cite{Karch2023}:
\begin{equation}\label{eq:upper}
c_{\text{eff}} \leq \min (c_L,c_R).
\end{equation}
In fact, this is consistent with the upper bound $c_{LR} \leq \min (c_L,c_R)$ shown in \cite{Quella2007, Meineri2020}.
Combining it with our conjectured inequality (\ref{eq:lower}), 
\begin{equation}
0\leq c_{LR} \leq c_{\text{eff}} \leq \min (c_L,c_R).
\label{eq:summ}
\end{equation}

Another result of this work is that these inequalities are sharp. For holographic CFTs, we were able to find the conditions under which interfaces saturate the bounds. 
We also show that $c_{LR} = c_{\text{eff}}$ is satisfied only if $c_{\text{eff}}$ is either $\min (c_L,c_R)$ or 0.
This means that the amount of energy transmission and information transmission across the interface match only in the case of the simplest interfaces, which are either boundaries or topological interfaces. 
We expect that these results will contribute to the understanding of non-topological interfaces.

\section{Holographic proof of the bound $c_{LR}\leq c_{\text{eff}}$ and its saturation}


The relation between $c_\text{eff}$ and the transmission coefficient $\mathcal{T}$ (or equivalently $c_{LR}$) in ICFT$_2$ has been studied in certain one-parameter families of conformal interfaces \cite{Sakai2008,Brehm2015}, where a monotonous function $c_\text{eff}(\mathcal{T}$) was found for free boson as well as certain lattice models. It is tempting to generalize this relation. However as we will show below, they are generally independent quantities. Instead, we prove that in a holographic ICFT$_2$, there is an inequality \eqref{eq:lower} between them. Moreover, the saturation of this bound in both holographic theories and free boson/fermion theories is realized when either $c_{LR}=c_\text{eff}=0$, or $c_{LR}=c_\text{eff}=c_L=c_R$.




\begin{figure}[t]
 \begin{center}
  \includegraphics[width=5.0cm,angle=90]{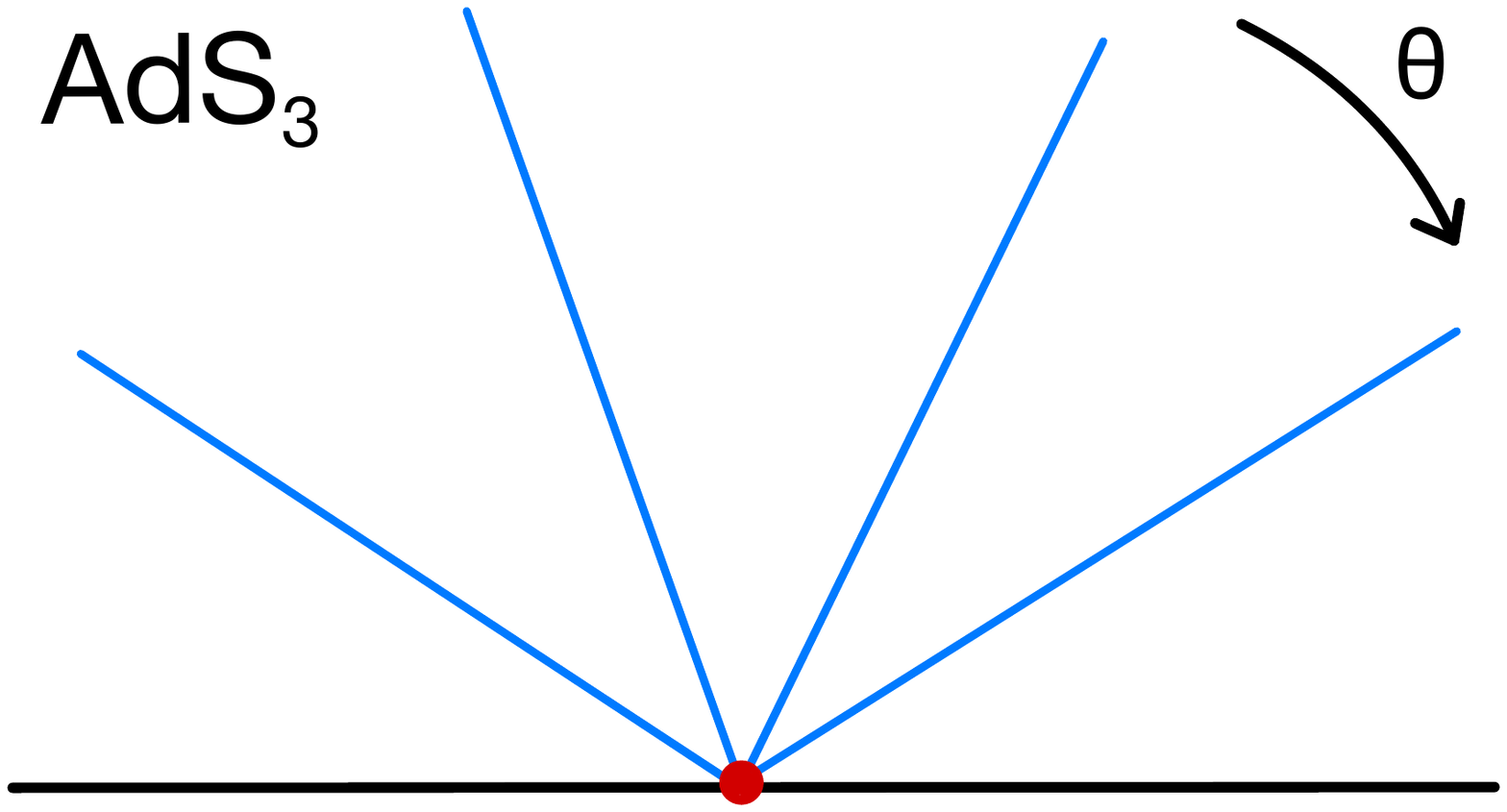}
 \end{center}
 \caption{The foliation of an asymptotic AdS$_3$ bulk. The black line below corresponds to the AdS asymptotic boundary, and the red dot is the one-dimensional interface. Each blue line represents an AdS${}_2$ slice, and in certain degeneration limit of the warp factor it can be a thin brane across which the effective AdS radius jumps.}
 \label{fig:ads}
\end{figure}

Consider bottom-up AdS/CFT where an ICFT$_2$ is dual to an asymptotic AdS$_3$ spacetime with sliced AdS$_2$ leaves and $SO(2,1)$ isometry\footnote{We do not consider (top-down) bulk configurations with dependence on extra dimensions in this article (e.g., SUSY Janus).}, as shown in Figure \ref{fig:ads}. Its metric is 
\begin{equation}
    ds^2=a^2(\theta)\left(\frac{dx^2-dt^2}{x^2}+d\theta^2\right),
    \label{eq:holobulk}
\end{equation}
where the AdS$_2$ is written in Poincaré patch, and $\theta\in (-\pi/2,\pi/2)$ is the slicing coordinate. $a(\theta)$ is a general function, referred to as the {\it warp factor}. 


Below, we will consider a general warp factor $a(\theta)$ that is a $C^2$ function. Near $\theta\rightarrow\pm\pi/2$, we have the asymptotic form $\lim_{\theta\rightarrow-\pi/2}a(\theta)= l_{L}/\cos\theta$ and $\lim_{\theta\rightarrow\pi/2}a(\theta)= l_{R}/\cos\theta$. Recall that the Brown-Henneaux (BH) formula relates $l_{L/R}$ to the central charges $c_{L/R}$ of CFT$_{L/R}$ as $c_{L/R}=3l_{L/R}/2G_N$ where $G_N$ is the Newton constant in 3D.

For such a continuous AdS domain wall solution, the transmission coefficient of the interface (or equivalently, $c_{LR}$ defined above) is given by \cite{Bachas2020}
\begin{equation}
    c_{LR}=\frac{3}{G_N}\left(\frac{1}{l_L}+\frac{1}{l_R}+8\pi G_N\sigma\right)^{-1},
\end{equation}
where $\sigma$ characterizes the net brane tension. To calculate the effective net tension, we first define a function $L(\theta)$ as
\begin{equation}
    L(\theta)\equiv \frac{a(\theta)}{\sqrt{1+\left(\frac{a'(\theta)}{a(\theta)}\right)^2}}.
    \label{eq:Ltheta}
\end{equation}
$L(\theta)$ represents an effective local AdS curvature radius.



\noindent
To calculate the brane tension $\sigma$ we note that the Isreal junction condition \cite{Israel1966, Lanczos1924} gives the differential change in brane tension needed to support the change in curvature radius \cite{Bachas_2023}
\begin{equation}
    8\pi G_N\frac{d\sigma}{d\theta}=\frac{a(\theta)|L'(\theta)|}{L(\theta)^2\sqrt{a(\theta)^2-L(\theta)^2}}.
    \label{eq:dsigma}
\end{equation}


Set $L_{j}$ to be the set of solutions for $L'(\theta)=0$ in $\theta\in(-\pi/2,\pi/2)$ where $j=1,\dots,M$. The integration is bounded by\footnote{The set of solutions can be empty, in which case there is no $L_{j}$, or it can be a collection of intervals, in which case we choose $L_{j}$ to be the right bound of the intervals. None of these cases affect our results.}
\begin{equation}
\begin{aligned}
    &8\pi G_N\int_{-\pi/2}^{\pi/2}d\theta\frac{d\sigma}{d\theta}\ge \int_{-\pi/2}^{\pi/2}d\theta\frac{|L'(\theta)|}{L^2(\theta)} \\
    &=\int |dL|\frac{1}{L^2}=\left|\frac{1}{l_L}-\frac{1}{L_{1}}\right|+\cdots+\left|\frac{1}{L_{M}}-\frac{1}{l_R}\right|.
\end{aligned}
\label{eq:ineqcont}
\end{equation}

The total net tension $\sigma$ can be derived by integrating $d\sigma/d\theta$ over its $C^2$ support of $\theta$ \cite{Bachas2021,Baig_2022,Bachas_2023}. Let the global minimum of $L(\theta)$ function be $l_\text{min}$. By definition, $l_\text{min}$ is equal to one of the $L_{j}$. Picking out $\theta_{\text{min}}$ in \eqref{eq:ineqcont} means that
\begin{equation}
    8\pi G_N\sigma\ge\left|\frac{1}{l_L}-\frac{1}{l_{\text{min}}}\right|+\left|\frac{1}{l_R}-\frac{1}{l_{\text{min}}}\right|.
\end{equation}
This leads us to conclude that 
\begin{equation}
\begin{aligned}
    c_{LR}&=\frac{3}{G_N}\left(\frac{1}{l_L}+\frac{1}{l_R}+8\pi G\sigma\right)^{-1} \\
    &\le\frac{3}{G_N}\left(\frac{1}{l_L}+\frac{1}{l_R}+\left|\frac{1}{l_L}-\frac{1}{l_{\text{min}}}\right|+\left|\frac{1}{l_R}-\frac{1}{l_{\text{min}}}\right|\right)^{-1}\\
    &\leq \frac{3}{G_N}\left(\frac{2}{l_\text{min}}\right)^{-1}\le \frac{3a_{\text{min}}}{2G_N}=c_\text{eff},
\end{aligned}
\end{equation}
where the second line is from \eqref{eq:Ltheta} and recalling the universal formula for the effective central charge in holographic ICFTs \cite{karch2021,Karchw_2023}. 

In order to saturate this inequality, from \eqref{eq:ineqcont}, the warp factor has to diverge wherever $|L'(\theta)|>0$. Hence there are two ways to realize $c_{LR}=c_\text{eff}$. One is when $a_\text{min}=l_\text{min}=0$ and the net brane tension diverges. In this case, $c_{LR}=c_\text{eff}=0$, and the two BCFTs are uncorrelated at all. The other is when $L(\theta)$ is constant, and the ICFT$_2$ is dual to a pure AdS$_3$ with a topological interface. In particular $c_{LR}=c_\text{eff}=c_L=c_R$. 

It is worth mentioning that holographic duals with discontinuity in $a'(\theta)$ are often considered as thin branes anchoring on the AdS boundary \cite{Bachas2020,Baig_2022}. It corresponds to a delta function in \eqref{eq:dsigma}. Upon integrating, it contributes to the net tension a term $8\pi G_N\sigma_t$ that follows the Coleman–De Luccia bound \cite{PhysRevD.21.3305}
\begin{equation}
    8\pi G_N\sigma_t\ge \left|\frac{1}{L_\text{left}}-\frac{1}{L_\text{right}}\right|
\end{equation}
where $L_\text{left}$ and $L_\text{right}$ are the effective AdS$_2$ radii on the left and right of the thin brane, respectively. The equality holds only when the AdS$_2$ radius diverges at the brane. It is obvious that our proof follows through in this degenerate limit of the warp factor, and so does the saturation condition.

As a corollary of the above proof, the transmission coefficient $c_{LR}$ depends on an integration of functions on the warp factor $a(\theta)$ over the entire range, while the entanglement entropy $c_\text{eff}$ only depends on the minimal value of $a(\theta)$. Therefore, in general, there is no strict monotonicity (correlation) between those two quantities.


The free boson/fermion theories with permeable interfaces provide another evidence for the inequality.  In both cases, there is a parameter $s\in [0,1]$ controlling the jumping radii on the two sides that characterize the interface, and the transmission coefficient is $\mathcal{T}=s^2$. 

\begin{description}

\item[$c=1$ free boson]  
The entanglement entropy for $c=1$ free theories across the interface has been derived to be \cite{Sakai2008}
\begin{equation}
\begin{aligned}
    c_\text{eff}^\text{bos}=&\frac{1}{2}+s+\frac{3}{\pi^2}\left((s + 1) \log (s + 1) \log s    \right. \\
    &\left. + (s - 1)~ \text{Li}_2(1 - s) + (s + 1)~\text{Li}_2(-s)\right),
\end{aligned}
\end{equation}
where $\text{Li}_2(s)$ is the dilogarithm function.
Arithmatically, we always have $c_\text{eff}^\text{bos}\ge c_{LR}^\text{bos}=s^2$ and the equality saturates only when $\mathcal{T}=s^2=0,1$.

\item[$c=1/2$ free fermion] 
The entanglement entropy for free fermion is \cite{Brehm2015}
\begin{equation}\label{eq:fermion}
\begin{aligned}
    c_\text{eff}^\text{fer}=&\frac{s-1}{2}-\frac{3}{\pi^2}\left((s + 1) \log (s + 1) \log s  \right. \\ 
    &\left. + (s - 1)~ \text{Li}_2(1 - s) + (s + 1)~\text{Li}_2(-s)\right).
\end{aligned}
\end{equation}

Again we have $c_\text{eff}^\text{fer}\ge c_{LR}^\text{fer}=s^2/2$, and the equality holds iff $\mathcal{T}=s^2=0,1$.

\end{description}
It is obvious that the free theories also saturate this bound only when $c_{LR}=c_{\text{eff}}$ is at either end of their spectrum. Therefore, we propose that this saturation condition for $c_{LR}\le c_{\text{eff}}$ is a universal feature among all ICFT$_2$.

Another evidence for the inequality comes from the defect perturbation.
Consider deforming a topological defect on a line $\gamma$ by a relevant or marginal defect field $\phi$,
\begin{equation}
\delta S = \lambda \int_\gamma dw \phi(w),
\end{equation}
where $\lambda$ is the coupling.
Under this perturbation, the effective central charge changes as follows up to order $\lambda^2$ \cite{Brehm2021},
\begin{equation}\label{eq:perturbative}
c_{\text{eff}} = c \pa{\ca{T}+\fr{1}{4}\ca{R}}.
\end{equation}
Note that this is consistent with (\ref{eq:fermion}), and the point is that the result (\ref{eq:perturbative}) is not limited to free fermion but holds in general.
Since $c\ca{T}=c_{LR}$ and $\ca{R} \geq 0$, we obtain
\begin{equation}
 c_{LR} \leq c_{\text{eff}}.
\end{equation}

\section{Holographic Saturation of $c_{\text{eff}}\le \min\{c_L,c_R\}$}


The upper bound \eqref{eq:upper} on $c_\text{eff}$ has been derived for both holographic theories and general ICFT$_2$ in \cite{Karch2023}. Below we will write down its saturation condition in holographic theories in terms of conditions on the {\it warp factor}, which is much more mathematically tractable compared to the CFT side. In particular, we show that with the possible presence of thin branes in the bulk, there is a much broader family of holographic ICFT$_2$ that saturates this bound than ICFT$_2$ with topological (transparent) interfaces.

For a holographic ICFT$_2$ dual to the bulk \eqref{eq:holobulk} with warp factor $a(\theta)$, we construct an auxiliary function \cite{Karch2023} 
\begin{equation}
    F=\frac{1}{L^2}=\left(\frac{a'}{a^2}\right)^2+\frac{1}{a^2}.
\end{equation}
The derivative of $F$ gives
\begin{equation}
    F'=\frac{2a'(a''a-2a'^2-a^2)}{a^5}.
    \label{eq:df}
\end{equation}
The null-energy condition (NEC) on $a(\theta)$ reads
\begin{equation}
    a^2(\theta)+2a'^2(\theta)-a(\theta)~a''(\theta)\geq 0.
    \label{eq:nec}
\end{equation}
In addition, from the achronal-averaged null energy condition (AANEC), minimum of $a(\theta)$ and maximum of $F(\theta)$ is not reached at thin branes \cite{Baig_2022}. Therefore, the minimal value of $a(\theta)$ at $\theta_{\text min}$ corresponds to the maximal value of $F(\theta)$\footnote{Its existence can be seen from \eqref{eq:df} and \eqref{eq:nec}.}. Concretely, we have 
\begin{equation}
   F_{\text max}=\left(\frac{3}{2G_N}\right)^2 \frac{1}{c_\text{eff}^2},\quad \lim_{\theta\rightarrow\pm\pi/2} F= \left(\frac{3}{2G_N}\right)^2 \frac{1}{c_{R/L}^2}. 
\end{equation}
If we set $c_L\ge c_R$, the saturation of $c_{\text{eff}}\le \min\{c_L,c_R\}$ is then equivalent to: 
\begin{description}

\item[case (a) $c_L>c_R$]
Saturation of NEC, i.e., $a(\theta)=l_R/\cos\theta$, for $\theta\in (\theta_{\text min},\pi/2)$, and \textit{any} $a(\theta)$ for the rest of the region subject to Einstein's equation and boundary conditions, with the possible presence of thin branes.

\item[case (b) $c_L=c_R$]
Pure AdS$_3$ with a topological interface, or at least two minima for the warp factor at $\theta_{\text min1}<\theta_{\text min2}$. Saturation of NEC, i.e., $a(\theta)=l/\cos\theta$ where $l_L=l_R=l$, for $\theta\in (-\pi/2,\theta_{\text min1})\cup(\theta_{\text min2},\pi/2)$, and \textit{any} $a(\theta)$ for $(\theta_{\text min1},\theta_{\text min2})$ subject to Einstein's equation, with the possible presence of thin branes.

\end{description}


\section{Discussion}

Our results inspire various future works and applications.

\begin{itemize}
\item
We have proposed a universal relationship between entanglement through the interface and energy transmission, based on the AdS/CFT correspondence.
It is desirable to have a proof for general CFTs.
In \cite{Karch2023}, we were able to give a general proof of the upper bound on the effective central charge using the entropic $c$-theorem.
A similar approach might be useful for our current purpose as well.
It may also be possible to verify our results numerically using the lattice realization of the conformal interface \cite{Cogburn2023, Tang:2023chv}.

\item
We have identified holographic interfaces which saturate the bounds. It is important to determine the saturation condition for general CFTs. In fact, there are non-holographic CFTs which saturate the bounds, as we show with explicit examples in Appendix \ref{app:example}.

\item
It is also desirable to generalize our results to higher-dimensional CFTs. Two potential challenges in higher dimensions are 
the lack of the Virasoro symmetry and the growth of entanglement, which makes well-known numerical calculation methods like DMRG unusable. 

\item
The effective central charge plays an important role in the weak measurement and the pseudo entropy  (see, for example, \cite{Nishioka2021, Kanda2023, Sun2023}). The relationship revealed in our work can be useful in elucidating the properties of such quantities.

\end{itemize}

What draws our attention here is the observation that many analytical methods on the gravity side do not depend on the dimension $d$.
The successful generalization of the concept of the effective central charge to higher dimensions has been achieved using the AdS/CFT correspondence \cite{Karch2023}.
This is precisely because the calculation on the gravity side is not dependent on the dimension $d$.
Based on this insight, 
it is a very interesting challenge to predict how the results revealed in this article would change in higher dimensions using the AdS/CFT correspondence. Additionally, providing proof within CFT for such predictions is also an important challenge. 




\section*{Acknowledgments}
We would like to thank Costas Bachas, Ilka Brunner, Shinsei Ryu, and Yifan Wang for careful reading and valuable comments on a draft of this paper.
AK, HS, and MW are supported in part by the U.S. Department of Energy under Grant No. DE-SC0022021 and a grant from the Simons Foundation (Grant 651440, AK).
HO and YK are supported in part by the U.S. Department of Energy, Office of Science, Office of High Energy Physics, under Award Number DE-SC0011632. In addition,
YK is supported by the Brinson Prize Fellowship at Caltech and the INAMORI Frontier Program at Kyushu University.
HO is  supported in part by the Simons Investigator Award (MP-SIP-00005259), the Guggenheim Fellowship, the World Premier International Research Center Initiative, MEXT, Japan, and
JSPS Grants-in-Aid for Scientific Research 23K03379. 
This work was performed in part at
the Aspen Center for Physics, which is supported by NSF grant PHY-1607611, and at the Kavli Institute for Theoretical Physics (KITP), which is supported by NSF grant PHY-2309135.

\begin{widetext}

\appendix
\section{Exapmle of $c_{\text{eff}}$ Bound Saturation in CFT}\label{app:example}
In this section, we give an explicit example of an interface between two {\it different} CFTs that preserves entanglement.
Interfaces saturate the bound (\ref{eq:upper}) can be easily realized as thin-branes in the holographic setup.
However, it is not trivial whether such interfaces can be realized on CFT.
On this background, it would be valuable to show an explicit example of such interfaces in WZW model.\footnote{One may wonder if such an interface can be constructed as a product of a Verlinde line and a Cardy boundary. But in general, this cannot be a consistent interface in our setup because the theory $\ca{A}(G)$ is non-factorizable into $\ca{A}(G/P)$ and $\ca{A}(P)$.}

\subsection{Construction}

Consider a WZW model with global symmetry $G$ and choose a subgroup $P$ of $G$.
One non-trivial rational interface may be constructed by breaking the chiral algebra\footnote{It is one of the extensions of the Virasoro algebra: current algebra, $\mathcal{W}$-algebra, etc.} of the CFT $\ca{A}(G)$ to $\ca{A}(G/P) \oplus \ca{A}(P)$, as considered in \cite{Quella2002}.
We construct a rational interface in a similar way to \cite{Quella2002} following their notation.
Let us start with a charge conjugated theory,
\begin{equation}
\ca{H}^G = \bigoplus_{\mu \in \text{Rep}(\ca{A}(G))}  \ca{H}^G_\mu \otimes \bar{H}^G_{\mu^+},
\end{equation}
where $\ca{H}^G_\mu$ is the space of the states in the Verma module with respect to the chiral algebra $\ca{A}(G)$ labeled by $\mu$.
In the following, we abbreviate the set of the irreducible representations $\text{Rep}(\ca{A}(G))$ by $\text{Rep}(G)$.
For our purpose, it is convenient to decompose the irreducible representations of $\ca{A}(G)$ to those of $\ca{A}(G/P) \oplus \ca{A}(P)$,
\begin{equation}
\ca{H}^G_\mu = \bigoplus_{(\mu,a) \in \text{All}(G/P)} \ca{H}^{G/P}_{(\mu,a)} \otimes \ca{H}^P_a,
\end{equation}
where we define the set of coset labels allowed by the branching selection rule as
\begin{equation}\label{eq:branching}
\text{All}(G/P) = \{     (\mu,a) | \ca{P}\mu -a \in \ca{P}Q  \} \subset  \text{Rep}(G) \times \text{Rep}(P).
\end{equation}
We denote the root lattice associated with $G$ by $Q$,
and $\ca{P}$ is a projector from the weight lattice of $G$ to that of $P$.
For later use, we describe the set of allowed coset labels in another way as
\begin{equation}
\text{All}(G/P) = \{     (\mu,a) | Q_J(\mu) = Q_{J'}(a) \text{ for all } (J,J') \in \ca{G}_{\text{id}}  \},
\end{equation}
where we define the monodromy charges with respect to a simple current $J \in \text{Rep}(G)$ and $J' \in \text{Rep}(P)$ by
\begin{equation}
Q_J(\nu) = h_J + h_\nu -h_{J\nu} \ \ \ \ \ \text{mod } 1, 
\end{equation}
and define $\ca{G}_{\text{id}} $ by the abelian group of all pairs $(J,J')$ satisfying the following condition,
\begin{equation}
Q_J(\mu) = Q_{J'}(\ca{P}\mu).
\end{equation}
Using this decomposition, the partition function can be re-expressed as
\begin{equation}
Z=  \abs{  \sum_{(\mu,a) \in \text{All}(G/P)} \chi^{G/P}_{(\mu,a)}(\tau) \chi^P_a(\tau)    }^2,
\end{equation}
where $\chi$ is the character of the chiral algebra.

Let us consider a rational interface between the $\ca{A}(G)$-WZW model and the $\ca{A}(G/P)$-WZW model.
We propose the rational interface as
\begin{equation}\label{eq:IAB}
\ca{I}^{G \to G/P}_\rho = \sum_{(\mu,a) \in \text{All} (G/P)  }   d^{(u,a)}_\rho 
||(\mu,a);G/P|| \otimes \kett{a;P} ,
\end{equation}
where the coefficients $d^{(u,a)}_\rho$ are determined by the modular $S$-matrices of $\ca{A}(G)$ and $\ca{A}(P)$ as
\begin{equation}
d^{(u,a)}_\rho \equiv \fr{ S^{G}_{\mu \rho}   }{   S^{G}_{\mu 0}   } \fr{1}{\sqrt{ S^P_{a0} }    }.
\end{equation}
The Ishibashi state is given by
\begin{equation}
\ket{i}\rangle \equiv \sum_{N} \ket{i;N} \otimes   U \overline{\ket{i;N}},
\end{equation}
where $\ket{i;N}$ is a state in the Verma module $i$ labeled by $N$, and $U$ is an anti-unitary operator.
We define the Ishibashi-type projector between two representations with some label $i,j$ as
\begin{equation}
||a|| : {\ca{H}_a \otimes \ca{\bar{H}}_a}^{(i)} \to  {\ca{H}_a \otimes \ca{\bar{H}}_a}^{(j)}.
\end{equation}
This projector commutes with the chiral algebra generators $J_n^P$,
\begin{equation}
J_n^P ||a|| = ||a|| J_n^P.
\end{equation}

\subsection{Open string partition function}

One requirement for the interface is that the multiplicity of states in the interface Hilbert space $\mathcal{H}^I$ is a positive integer.
To check for our interface to be consistent, let us consider the partition function with two interfaces,
\begin{equation}
\begin{aligned}
Z &= \text{tr} \ \ex{-\fr{\beta}{2} H^{(G)}}
\ca{I}^{G \to G/P}_\rho
 \ex{-\fr{\beta}{2} H^{(G/P)}}
\ca{I}^{G/P \to G.}_\sigma \\
&=
 \sum_{(\mu,a) \in \text{All}(G/P)} 
d^{(u,a)}_\rho
\overline{d^{(u,a)}_\sigma}
\abs{ \chi^{G/P}_{(\mu,a)}(\tau )  }^2   \chi^P_a(\tau),
\end{aligned}
\end{equation}
where $H^{(G)}$ is the Hamiltonian of the theory $\ca{A}(G)$.
The dual-channel expansion can be obtained by using the modular-$S$ transformation as
\begin{equation}
\begin{aligned}
Z&= 
\sum_{\substack{    (\mu,a) \in \text{All}(G/P) \\  (\nu,b), (\bar{\nu},\bar{b} ) \in \text{Rep}(G/P) \\ c \in \text{Rep}(P)    }}
 \fr{ S^{G}_{\mu \rho}   }{   S^{G}_{\mu 0}   }
 \fr{ \bar{S}^{G}_{\mu \sigma}   }{   \bar{S}^{G}_{\mu 0}   }
 \fr{1}{S^P_{a0}}
 S^{G/P}_{(\mu,a), (\nu,b)}
  \bar{S}^{G/P}_{(\mu,a), (\bar{\nu},\bar{b})}
  S^P_{ac}
  \chi^{G/P}_{(\nu,b)}\pa{- \fr{1}{\tau}}
  \chi^{G/P}_{(\bar{\nu},\bar{b})}\pa{- \fr{1}{\tau}}
  \chi^P_c \pa{- \fr{1}{\tau}}\\
&=
\abs{\ca{G}_{id}}^2
\sum_{\substack{   \eta \in \text{Rep}(G) \\   (\mu,a) \in \text{All}(G/P) \\  (\nu,b), (\bar{\nu},\bar{b} ) \in \text{Rep}(G/P) \\ c \in \text{Rep}(P)    }}
{N_{\rho \sigma^+}}^\eta
\fr{S^G_{\mu \eta}  S^G_{\mu \nu} \bar{S}^G_{\mu \bar{\nu}} }{S^G_{\mu 0}}
\fr{S^P_{a\bar{b}}S^P_{ac} \bar{S}^P_{ab}  }{S^P_{a0}}
  \chi^{G/P}_{(\nu,b)}\pa{- \fr{1}{\tau}}
  \chi^{G/P}_{(\bar{\nu},\bar{b})}\pa{- \fr{1}{\tau}}
  \chi^P_c \pa{- \fr{1}{\tau}}.
 \end{aligned}
\end{equation}
In the second step, we use the following relation,
\begin{equation}
S^{G/P}_{(\mu,a), (\nu,b)} = \abs{\ca{G}_{id}} S^G_{\mu \nu} \bar{S}^P_{ab},
\end{equation}
and the fusion rule of the quantum dimensions,
\begin{equation}\label{eq:qd}
 \fr{ S^{G}_{\mu \rho}   }{   S^{G}_{\mu 0}   }  \fr{ \bar{S}^{G}_{\mu \sigma}   }{   \bar{S}^{G}_{\mu 0}   }
 =
 \sum_{\eta \in \text{Rep}(G)}
 {N_{\rho \sigma^+}}^\eta  \fr{ S^{G}_{\mu \eta}   }{   S^{G}_{\mu 0}   },
\end{equation}
where $ {N_{\rho \sigma^+}}^\eta $ is the fusion coefficient.
It is not so easy to take the summation over $\mu$ and $a$ because the pairs $(\mu, a)$ follow the branching selection rule (\ref{eq:branching}).
One can remove this restriction by inserting the projection operator,
\begin{equation}
P(\mu,a) = \fr{1}{\abs{\ca{G}_{id} } } \sum_{(J,J') \in \ca{G}_{id}} \ex{2\pi i \pa{Q_J(\mu) - Q_{J'}(a)}  }.
\end{equation}
The resulting expression is
\begin{equation}
\begin{aligned}
Z&=
\abs{\ca{G}_{id}}^2
\sum_{\substack{  \mu, \eta \in \text{Rep}(G) \\ (\nu,b), (\bar{\nu},\bar{b} ) \in \text{Rep}(G/P) \\ a,c \in \text{Rep}(P)    }}
P(\mu,a)
{N_{\rho \sigma^+}}^\eta
\fr{S^G_{\mu \eta}  S^G_{\mu \nu} \bar{S}^G_{\mu \bar{\nu}} }{S^G_{\mu 0}}
\fr{S^P_{a\bar{b}}S^P_{ac} \bar{S}^P_{ab}  }{S^P_{a0}}
  \chi^{G/P}_{(\nu,b)}\pa{- \fr{1}{\tau}}
  \chi^{G/P}_{(\bar{\nu},\bar{b})}\pa{- \fr{1}{\tau}}
  \chi^P_c \pa{- \fr{1}{\tau}}  \\
&=
\abs{\ca{G}_{id}}
\sum_{\substack{  \mu, \eta \in \text{Rep}(G) \\ (\nu,b), (\bar{\nu},\bar{b} ) \in \text{Rep}(G/P) \\ a,c \in \text{Rep}(P) \\ (J,J') \in \ca{G}_{id}    }}
{N_{\rho \sigma^+}}^\eta
\fr{S^G_{\mu \eta}  S^G_{\mu, J\nu} \bar{S}^G_{\mu\bar{\nu}} }{S^G_{\mu 0}}
\fr{S^P_{a\bar{b}}S^P_{ac} \bar{S}^P_{a, J'b}  }{S^P_{a0}}
  \chi^{G/P}_{(\nu,b)}\pa{- \fr{1}{\tau}}
  \chi^{G/P}_{(\bar{\nu},\bar{b})}\pa{- \fr{1}{\tau}}
  \chi^P_c \pa{- \fr{1}{\tau}} \\
&=
\abs{\ca{G}_{id}}
\sum_{\substack{  \eta \in \text{Rep}(G) \\  (\nu,b) \in \text{All}(G/P)  \\ (\bar{\nu},\bar{b} ) \in \text{Rep}(G/P) \\ c \in \text{Rep}(P)    }}
{N_{\rho \sigma^+}}^\eta
{N_{\eta \nu}}^{\bar{\nu}}
{N_{\bar{b}c}}^b
  \chi^{G/P}_{(\nu,b)}\pa{- \fr{1}{\tau}}
  \chi^{G/P}_{(\bar{\nu},\bar{b})}\pa{- \fr{1}{\tau}}
  \chi^P_c \pa{- \fr{1}{\tau}}.
\end{aligned}
\end{equation}
In the second step,
we use the transformation law of the modular-$S$ matrix,
\begin{equation}
S^G_{J \mu, \nu} = \ex{2\pi i Q_J(\nu)} S^G_{\mu \nu}.
\end{equation}
In the last step, we use the invariance of the character $\chi^{G/P}_{(\nu,b)} = \chi^{G/P}_{(J\nu, J'b)}$ for $(J,J') \in \ca{G}_{id}$, re-label $(J\nu, J'b)$ by $(\nu,b)$,
and apply the Verlinde formula,
\begin{equation}
{N_{\mu \nu}}^\rho = \sum_{\sigma \in \text{Rep}(\ca{A})} \fr{ S_{\sigma \mu} S_{\sigma \nu} \bar{S}_{\sigma \rho}  }{S_{\sigma 0}}.
\end{equation}
The last expression implies that the multiplicity of the states in the interface Hilbert space is a positive integer.

\subsection{Effective Central Charge}

To evaluate the effective central charge, we consider the following replica partition function,
\begin{equation}
Z_n=
 \text{tr} \ \pa{\ex{-\fr{\beta}{2} H^{(G)}}
 \ca{I}^{G \to G/P}_\rho
 \ex{-\fr{\beta}{2} H^{(G/P)}}
\ca{I}^{G/P \to G.}_\rho 
}^n.
\end{equation}
In the same way as the double-interface case,
the multiple-interface Hilbert space has positive integer multiplicities,
which can be shown by using the fusion rule of the quantum dimensions (\ref{eq:qd}).
Consequently,
the replica partition function can be calculated in the $\beta \to 0$ limit as
\begin{equation}\label{eq:E0}
Z_n \propto \ex{\fr{c^{(G/P)}}{12} \fr{1}{n\beta}+\fr{c^{(P)}}{24}\fr{n}{\beta}},
\end{equation}
where $c^{(G/P)}$ and $c^{(P)}$ are the central charges of the theory $\ca{A}(G/P)$ and $\ca{A}(P)$.
Following the definition (\ref{eq:ceff}), we obtain the effective central charge for the interface between the theory $\ca{A}(G)$ and $\ca{A}(G/P)$ as
\begin{equation}
c_{\text{eff}}= c^{(G/P)} = \min (c^{(G/P)}, c^{(G)}).
\end{equation}
We can see that the interface (\ref{eq:IAB}) is an explicit realization of the saturation of the upper bound (\ref{eq:upper}).

\end{widetext}

\bibliographystyle{JHEP}
\bibliography{main}

\end{document}